\documentclass[journal,twocolumn]{IEEEtran}
\usepackage{epsfig,makeidx,color,epstopdf}
\usepackage{amsmath,amssymb,amsthm,bbm}
\usepackage{cite,graphicx}
\usepackage{enumerate}
\def\indicator{{\mathbbmtt 1}}
\DeclareMathOperator*{\argmin}{\arg\!\min}
\DeclareMathOperator*{\argmax}{\arg\!\max}

\def\cK{{\cal K}}

\def\rH{{\rm H}}
\def\rT{{\rm T}}

\def\uE{{\mathbb E}}

\newtheorem{mytheorem}{\bf Theorem} % [section]
\def\deft{ \buildrel \triangle \over = }

\def\be{ \begin{equation} }
\def\ee{ \end{equation} }
\def\bea{ \begin{eqnarray} }
\def\eea{ \end{eqnarray} }

\def\by{{\bf y}}
\def\bc{{\bf c}}

\def\bs{{\bf s}}
\def\ba{{\bf a}}

\def\bn{{\bf n}}
\def\bk{{\bf k}}

\def\bA{{\bf A}}

\def\bC{{\bf C}}

\def\bI{{\bf I}}

\def\bN{{\bf N}}
\def\bS{{\bf S}}

\def\bR{{\bf R}}

\def\bY{{\bf Y}}

\def\b0{{\bf 0}}

\def\cC{{\cal C}}

\def\cR{{\cal R}}
\def\cI{{\cal I}}
\def\cN{{\cal N}}

\ifCLASSOPTIONonecolumn
  \newcommand{\figwidth}{0.50\columnwidth}
  
\else
  \newcommand{\figwidth}{0.90\columnwidth}
  
\fi

\IEEEoverridecommandlockouts

\begin{document}

\title{Compressive Random Access 
with Multiple Resource Blocks and Fast Retrial}

\author{Jinho Choi\\
\thanks{The author is with
the School of Information Technology,
Deakin University, Geelong, VIC 3220, Australia
(e-mail: jinho.choi@deakin.edu.au).}}

\date{today}
\maketitle

\begin{abstract}
In this paper, we propose a compressive random access (CRA) scheme 
using multiple resource blocks (RBs) to support
massive connections for machine type communications (MTC).
The proposed CRA scheme is scalable. As a result, if the
number of devices increases, more RBs can be added to support them.
Thanks to multiple RBs, we 
can employ fast retrial between RBs for re-transmissions of collided packets, 
which can result in short access delay.
For stable CRA with fast retrial,
we derive conditions (with a rate control scheme),
and analyze the steady state performance to find the throughput and 
delay. 
Through analysis and simulation results,
we can see that the proposed scheme can perform better than
conventional multichannel ALOHA and enjoy a trade-off between
the performance and complexity in terms of the number of RBs.
\end{abstract}


\ifCLASSOPTIONonecolumn
\baselineskip 26pt
\fi

\section{Introduction}

In 5th generation (5G) systems,
there has been a growing interest in
machine-type communications (MTC)
for the Internet of Things (IoT) \cite{Hasan13} \cite{Shari15} 
\cite{Bockelmann16}.
The applications of MTC are diverse from health care
to smart grid,
where a huge number of devices exist in the system, but
only a few of them are active at a particular timing instance.
Therefore, random access is suitable for MTC to accommodate
a number of devices with sparse
activity  \cite{Lee11, Hasan13, Niyato14}.

A random access scheme, called random access
(RACH) procedure, has been proposed for
a cellular system, i.e., the long term evolution-advanced (LTE-A) system
\cite{3GPP_MTC}.
RACH procedure is a contention-based random access method,
which is similar to the slotted ALOHA protocol \cite{BertsekasBook}.
In RACH procedure, there are multiple preambles
in a preamble pool, 
and a device can select a preamble randomly from 
the pool and transmits the selected one for access. 
It can be connected if there is no collision, i.e., 
the preamble is not transmitted by any other devices.
In addition to RACH procedure, there is a similar approach 
in LTE-A for MTC, which is called narrowband-IoT (NB-IoT)
\cite{3GPP_NBIoT}.

Recently, the notion of compressive sensing (CS)
\cite{Candes05, Donoho06, Eldar12} is employed
to exploit the sparsity of active devices 
for multiuser detection (MUD) in random access
\cite{Zhu11} \cite{Schepker13} \cite{Wunder15A} \cite{Beyene17}
\cite{Liu18}.
Among many devices present in MTC, only a few of them 
attempt to access the network by transmitting their signature waveforms.
The sparse activity of devices, which can be modeled as a sparse vector,
allows the principle of 
CS to be effectively applied to
MUD with low complexity.
When CS based MUD is employed,
the resulting random access scheme is often called
compressive random access (CRA) \cite{Wunder15A}.
CRA is similar to code division multiple access (CDMA)
\cite{ViterbiBook} \cite{TseBook05}
in a way to transmit data symbols 
(where spreading codes are used to transmit data symbols with multiple
access), while 
it is a random access scheme that requires active user detection 
as the spreading codes used are unknown in advance at a receiver.
In CRA, spreading codes are not 
necessarily orthogonal
due to MUD that can detect multiple signals in the
presence of interference \cite{VerduBook} \cite{ChoiJBook}.
Consequently, the main advantage of CRA over conventional multichannel
random access schemes based on orthogonal channels (e.g.,
multichannel ALOHA) is the increase of the number of channels that
can reduce the probability of collision and improve the throughput
\cite{Choi17_Stability}
at the cost of increased complexity at a receiver.
In addition, CRA supports grant-free transmissions.
Thus, as in \cite{Liu18}, grant-free CRA can be more efficient
than grant-based schemes (e.g., \cite{3GPP_MTC}, \cite{3GPP_NBIoT})
since it does not need to wait for the grant from an access point (AP).

Grant-free CRA schemes \cite{Zhu11, Schepker13,
Wunder15A, Liu18} can be characterized by 
a unique signature sequence for each device.
Thus, the number of devices 
becomes limited by the number of signature sequences.
To support many devices in MTC, 
although a wide bandwidth is considered,
this can result in two difficulties: 
\emph{i)} a long sequence is required to accommodate a number of devices
regardless of sparse activity;
\emph{ii)}
the complexity of CS algorithms for MUD can be high due to 
a large number of columns (or signature vectors) in a measurement matrix
(a high computational complexity).
To avoid the above difficulties,
we can apply the approach used in the RACH procedure 
to grant-free CRA.
That is, instead of assigning a unique signature to each device,
a randomly selected signature 
from a pool of pre-determined signatures can be used when an active
device is to transmit signals to a receiver or AP,
where a signature code is used as a spreading code
to transmit data symbols as in CRA.
This approach is studied in \cite{Choi16a},
while it suffers from collision that
happens when multiple active devices choose the same signature.

In this paper, we generalize the CRA scheme in \cite{Choi16a}
using multiple resource blocks (RBs) 
in a multicarrier system with a certain re-transmission strategy. 
In the proposed scheme,
we divide subcarriers into multiple groups, or RBs,
so that it becomes scalable,
since more RBs can be added to support more devices in MTC.
At the AP, parallel multiple CS-based detectors can be employed 
for MUD with low complexity.
In addition, thanks to multiple RBs,
we can employ fast retrial \cite{YJChoi06} between RBs
for re-transmissions of collided packets.
In fast retrial, when a device experiences collision in the current
time slot, this device can re-transmit 
in the next time slot without random backoff. As a result,
the access delay can be short.
In summary,
the main advantage of the proposed CRA
scheme over other grant-free CRA schemes (with single big RB)
\cite{Zhu11, Wunder15A, Beyene17, Liu18}
is mainly three-fold: 
{\it i)} scalability;
{\it ii)} a low computational complexity for CS based MUD; 
{\it iii)} fast retrial between RBs for short access delay.  
The main contributions\footnote{Note that multiple 
RB based CRA has been studied in \cite{Choi16_GC}.
However, in \cite{Choi16_GC}, no re-transmission is considered
(this paper is an extension of \cite{Choi16_GC}).}
of the paper can also be summarized
as follows: {\it a)} a scalable grant-free CRA scheme is proposed
not only to support a large number of devices with a reasonably
low complexity for CS based MUD, but also to accommodate
fast retrial;
{\it b)} the stability and steady-state performance are analyzed
under certain assumptions to see the impact of
the number of RBs as well as other parameters
on the performance of the proposed CRA scheme.

The rest of the paper is organized as follows.
In Section~\ref{S:SM}, we present a system model for CRA
with multiple RBs in a multicarrier system.
We discuss CS based MUD in Section~\ref{S:CS}
and propose a model for its recovery performance in CRA.
With a rate control scheme, we analyze the stability of the proposed CRA
scheme with fast retrial in Section~\ref{S:SA}.
To understand the throughput and access delay,
we study a steady state analysis in Section~\ref{S:PA}.
In Section~\ref{S:Sim},
simulation results are presented with theoretical results
obtained from Section~\ref{S:PA}.
We finally conclude the paper with some remarks in Section~\ref{S:Conc}.

{\it Notation}:
Matrices and vectors are denoted by upper- and lower-case
boldface letters, respectively.
The superscripts $\rT$ and $\rH$
denote the transpose and complex conjugate, respectively.
%The $p$-norm of a vector $\ba$ is denoted by $|| \ba ||_p$
%(If $p = 2$, the norm is denoted by $||\ba||$ without the subscript). 
$\uE[\cdot]$
and ${\rm Var}(\cdot)$
denote the statistical expectation and variance, respectively.
$\cC\cN(\ba, \bR)$
represents the distribution of
circularly symmetric complex Gaussian (CSCG)
random vectors with mean vector $\ba$ and
covariance matrix $\bR$.

\section{System Model}	\label{S:SM}

\begin{figure}[thb]
\begin{center}
\includegraphics[width=\figwidth]{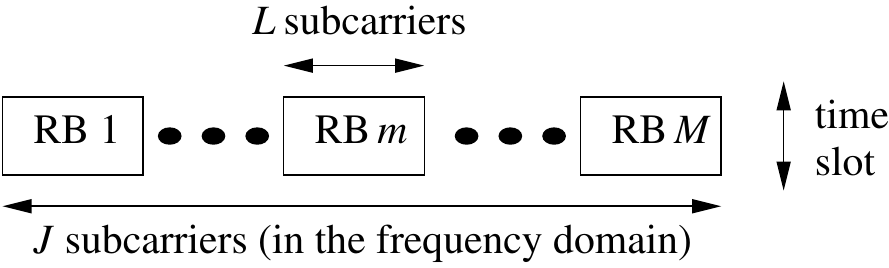}
\end{center}
\caption{An illustration of multiple RBs in the frequency
domain.}
        \label{Fig:RBs}
\end{figure}

Consider uplink transmissions from a number of devices to an AP.
Suppose that 
when a device becomes active to transmit a packet,
it can transmit its packet without any permission (i.e.,
grant-free transmissions are assumed) \cite{Liu18}.
For uplink transmissions,
a radio resource block is divided into
$M$ RBs and each RB is orthogonal to each other.
In each RB, there are $N$ spreading codes (SCs) of length $L$.
For grant-free CRA, an active device 
is to randomly choose an RB and an SC for data transmissions.
Throughout the paper, we assume a multicarrier system
that has $J = L M$ subcarriers. Thus, an RB
consists of $L$ subcarriers and each SC
is a multicarrier spread sequence of length $L$ \cite{FazelBook}.
In Fig.~\ref{Fig:RBs}, we illustrate multiple RBs
with $J$ subcarriers in the frequency domain.

Let $\bC = [\bc_0 \ \ldots \bc_{N-1}]$ denote
a matrix of $N$ SC vectors
where $\bc_n$ denotes the $n$th SC, which is a multicarrier 
spread sequence of length $L$.
Thus, the size of $\bC$ becomes $L \times N$, and
in general $L \le N$,
which implies that the SC sequences can be non-orthogonal.
For convenience, let
$$
\eta = \frac{N}{L},
$$
which is referred to as the virtual bandwidth expansion factor.
Note that if SC sequences are orthogonal to each other,
we have $\eta = 1$ (i.e., $N = L$).
We assume the same set of SC sequences for all RBs.

Suppose that
a packet consists of $T$ symbols and 
each active device can choose one of $M$
RBs uniformly at random.
At the AP, the received signal vector at (symbol) time $t$ over the $m$th RB
is given by
\begin{align}
\by_{m,t} & = [y_{m,t,0} \ \ldots \ y_{m,t,L-1}]^\rT \cr
& = \bC \bs_{m,t} + \bn_{m,t},
\ t = 0,\ldots,T-1,
	\label{EQ:yCsn}
\end{align}
where $y_{m,t,l}$ represents the received signal
at time $t$
through the $l$th subcarrier of the $m$th RB,
$\bs_{m,t}$ is a random access vector at time $t$, and
$\bn_{m,t} =  [n_{m,t,0} \ \ldots \ n_{m,t,L-1}]^\rT
\sim \cC \cN(0, N_0 \bI)$ is the background noise.
Note that the RBs are orthogonal. Thus, there is no interference
from the other RBs. 

Let $\cK_{m,n}$ denote
the index set of the active devices that
choose the $m$th RB and the $n$th SC.
Assuming that the channel gain remains
unchanged within the duration of a packet 
transmission, 
the $n$th element of $\bs_{m,t}$ is given by
\be
[\bs_{m,t}]_n = \sum_{k \in \cK_{m,n}} h_{k,m} x_{k,t},
	\label{EQ:smn}
\ee
where $h_{k,m}$ denotes the (frequency-domain) channel coefficient
from device $k$ to the AP over the $m$th RB and $x_{k,t}$ represents the 
signal from device $k$ at time $t$.
In \eqref{EQ:smn}, we assume that the bandwidth of an RB
is sufficiently narrow so that the frequency-domain channel gain
remains unchanged within the bandwidth of an RB.
The total number of active devices is given by
$K = \sum_{m=1}^M \sum_{n=1}^N |\cK_{m,n}|$
and the number of active devices that choose the $m$th RB is given by
$K_m =  \sum_{n=1}^N |\cK_{m,n}|$.
Thus, we have $K = \sum_{m=1}^M K_m$.
In general, we expect to have $K_m \ll N$
due to sparse activity.

For convenience,
the above random access
system with multiple RBs (i.e., $M > 1$) is referred to as the 
multiple RB (MRB) based CRA 
(MRB-CRA) system in this paper.
In addition, we employ fast retrial \cite{YJChoi06} between RBs
to re-transmit collided packets for short access delay.
That is, if an active device has a collided packet,
this packet can be immediately re-transmitted in the next time 
slot\footnote{In the paper, we assume that the length of packet
is equivalent to the length of time slot.}
through a randomly selected RB among $M$ RBs.

It is noteworthy that the resulting system
becomes scalable. Thus, if the number of devices increases,
we can add more RBs,
which of course increases the system bandwidth
as shown in Fig.~\ref{Fig:RBs}.
In addition, at an AP, there can be $M$ parallel MUDs
to recover sparse signals in each RB.
This feature might be important to reduce the processing time
at the AP since the total processing time becomes identical
to the processing time of MUD for one RB if a parallel processor
is used to implement the receiver \cite{Choi16_GC}.

Although we assume that 
an RB is selected uniformly at random in this paper,
it is also possible to choose it based on the channel
state information (CSI) at each active device to lower the transmission power.
In \eqref{EQ:smn}, we need to adjust 
the amplitude of $x_{k,t}$ to overcome
fading if the channel gain, $|h_{k,m}|$, is low.
In general, each device can choose the
amplitude of $x_{k,t}$ to be inversely proportional to
the channel gain to equalize the overall signal gain.
In this case, if $|h_{k,m}|$ is low,
the transmit power has to be high, which is not desirable.
In the MRB-CRA system, this problem
can be mitigated if an active device can choose
the RB of the highest channel gain.
That is, active device $k$ can choose the RB as follows:
$m(k) = \argmax_m |h_{k,m}|$,
where $m(k)$ denotes the index of the selected RB by active
device $k$. Since the channel gains are random,
the selection of RBs by active devices becomes also random
and active devices could be uniformly distributed over multiple
RBs in this case as well 
(provided that $h_{k,1}, \ldots, h_{k,M}$ are iid).
From this, throughout the paper, we assume
that the power control
is adopted to equalize
the overall signal gain as follows:
$$
|h_{k,m(k)} x_{k,t}| = \sqrt{P}, \ \mbox{for all active device $k$},
$$
where $P$ is the transmit power.
Thus, the signal to noise ratio (SNR) becomes ${\rm SNR} = \frac{P}{N_0}$.
Throughout the paper, 
we assume that the SNR is sufficiently high.
Note that a similar approach is considered in \cite{Stefanovic13},
where the phase of the signal is also compensated.

\section{CS based MUD}	\label{S:CS}

In this section, we focus on CS based MUD for one RB
(as each RB has the same structure)
to recover multiple signals and propose a model for the performance
analysis.

\subsection{CS based Signal Recovery}

For convenience, since we consider 
one RB, we omit the RB index $m$ if there is no risk of confusion.
Let $\bY = [\by_0 \ \ldots \ \by_{T-1}]$,
$\bS = [\bs_0 \ \ldots \ \bs_{T-1}]$,
and $\bN = [\bn_0 \ \ldots \ \bn_{T-1}]$.
Then, \eqref{EQ:yCsn} becomes
\be
\bY = \bC \bS + \bN.
	\label{EQ:bY}
\ee
Let $\cI$ denote the support of $\bs_t$.
If $\bC$ is seen as a measurement matrix,
the estimation of $\cI$ 
from $\bY$ in \eqref{EQ:bY} is a typical multiple measurement
vectors (MMV) problem \cite{Chen06} \cite{Davies12} as
the support of $\bs_t$ is the same for all $t$.
Under the high SNR assumption or ignoring $\bN$,
the MMV sparse recovery problem can be formulated as follows
\cite{Davies12}:
\be
\hat \bS = \argmin_\bS |{\rm supp} (\bS)| 
\ \mbox{subject to} \ \bC \bS = \bY,
	\label{EQ:MMVP}
\ee
where ${\rm supp} (\bS)$ represents the support of
matrix $\bS$, which is defined by
${\rm supp} (\bA) = \bigcup_i {\rm supp}(\ba_i)$.
Here, $\ba_i$ denotes the $i$th column vector of $\bA$.
In this case, 
a sufficient and necessary condition to estimate $\cI$
\cite{Chen06,Davies12}
is given by
\be
N < \frac{{\rm spark}(\bC) -1 + {\rm rank}(\bS)}{2},
	\label{EQ:Ms}
\ee
where $\operatorname{spark}(\bC)$ is
the smallest number of columns from  $\bC$ that are linearly dependent
\cite{Donoho03}.
It can be readily shown that
${\rm rank}(\bS) \le \min\{N, T\}$.
Since each element of $\bS$ is iid, 
we have
${\rm rank}(\bS) = N$ if $T \ge N$ with probability (w.p.) 1.
From this, \eqref{EQ:Ms} is reduced to 
\begin{align}
N & \le D
\deft {\rm spark}(\bC) - 2,
\end{align}
where $D$ is the sparsity threshold for a recovery guarantee.
If $\bC$ is random
(e.g., all the elements of $\bC$ are independent CSCG random variables)
and $K \ge L$,
${\rm spark}(\bC) -1 = {\rm rank}(\bC) = L$ w.p. 1
\cite{Bruckstein09}.
In this case, we have
\be
D = L - 1.
	\label{EQ:tL1}
\ee
However, in practice, the sparsity threshold 
might be smaller than 
$L-1$ due to various reasons. 
For example, for low-complexity implementations,
greedy algorithms 
can be used, which result in suboptimal performances.
In addition, due to the background noise, there might be performance
degradation.

\subsection{A Model for Recovery Performance}

Throughout the paper, we assume that 
the AP is able to recover the signals of up to $D$-sparsity in each RB,
where $D < L$. That is, we consider
the following assumptions.
\begin{itemize}
\item[{\bf A1}] If the number of active devices in an RB is greater than
$D$, the AP cannot recover them at all.
\item[{\bf A2}] If the number of active devices in an RB is less
than or equal to $D$, the AP recovers the active devices whose SCs
do not collide with each other.
\end{itemize}
Clearly, $D$ is a threshold value and
depends on the properties of $\bC$ including its
size, i.e., $N$ and $L$.
Note that although there are more than $D$ active devices,
the AP may be able to recover some of
them. However, for tractable analysis,
we simply assume {\bf A1}.
In {\bf A2}, it is assumed that
although the AP can recover all the signals,
it cannot resolve collided signals from the devices that choose
the same SC.
In general, in this paper, $D$ is seen as a performance 
indicator of CS based MUD to be estimated.

\begin{figure}[thb]
\begin{center}
\includegraphics[width=\figwidth]{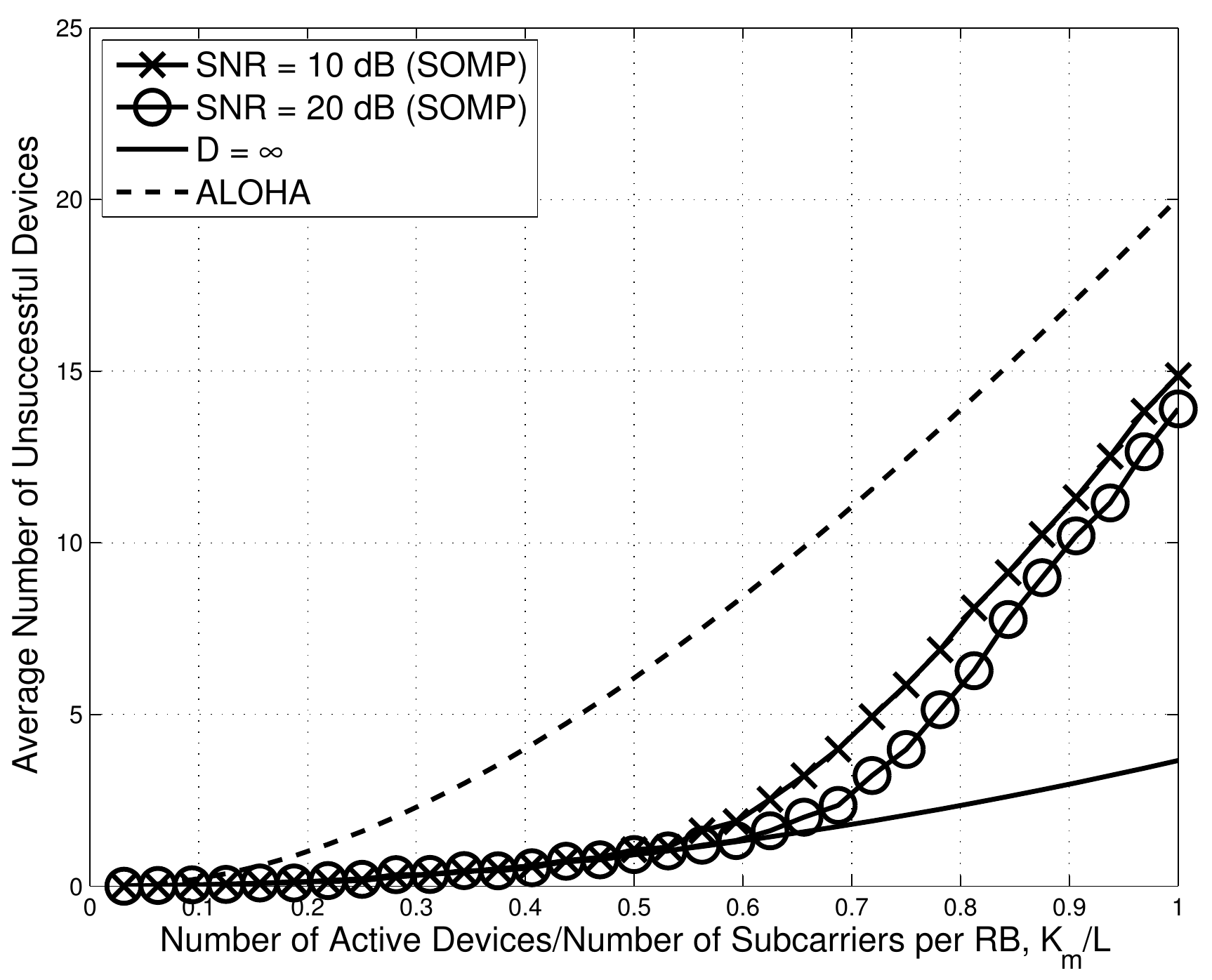} \\
(a) \\
\includegraphics[width=\figwidth]{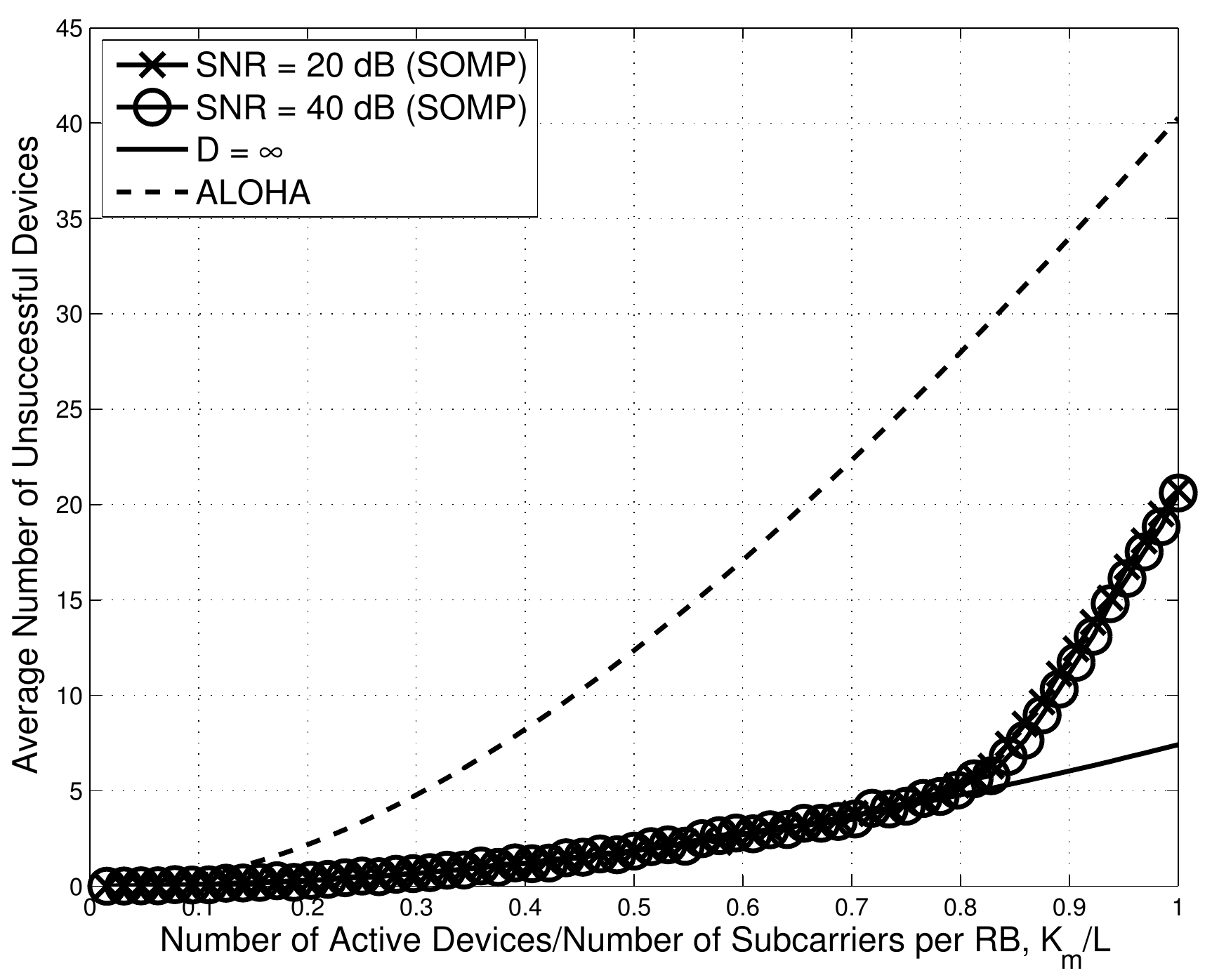} \\
(b) 
\end{center}
\caption{The average number of unsuccessful devices for 
different numbers of active devices per RB, $K_m$:
(a) $L = 32$, $T = 20L$, $N = 8L$, and  SNR $\in \{10, 20 {\rm dB}\}$;
(a) $L = 64$, $T = 20L$, $N = 8L$, and  SNR $\in \{20, 40 {\rm dB}\}$.}
        \label{Fig:Pplt}
\end{figure}

Based on the assumptions of {\bf A1} and {\bf A2},
when there are $K_m$ active devices in RB $m$,
since there are $N$ SCs per RB,
the average number of unsuccessful devices becomes
\be
U (K_m) = \left\{
\begin{array}{ll}
K_m - 
K_m \left(1 - \frac{1}{N} \right)^{K_m -1}, & \mbox{if $K_m \le D$;} \cr
K_m, & \mbox{if $K_m > D$.} \cr
\end{array}
\right.
	\label{EQ:UKm}
\ee
In Fig.~\ref{Fig:Pplt}, we show the average
number of unsuccessful devices, i.e., $U(K_m)$
when the simultaneous
orthogonal matching pursuit (S-OMP) algorithm proposed in \cite{Chen06}
\cite{Tropp06_SP}
is used for CS based MUD.
Here, for simulations, we assume that each element of $\bC$
is an independent CSCG random variable with zero mean and variance
$\frac{1}{L}$.
It is shown that $U(K_m)$ from the simulation results
follows $K_m - K_m 
\left(1 - \frac{1}{N} \right)^{K_m -1}$ when $K_m$ is small.
However, as $K_m$ approaches a certain value, 
$U(K_m)$ increases rapidly. Thus, the model 
in \eqref{EQ:UKm} can be used to 
analyze the performance of MRB-CRA,
where we assume that there exists
a threshold value $D$ such that $U(K_m) = K_m$ if $K_m > D$
(i.e., {\bf A1}). 

Note that
according to Figs.~\ref{Fig:Pplt} (a) and (b), $D$ might be dependent on
$L$. For example, 
when $L = 32$, as shown in Fig.~\ref{Fig:Pplt} (a), 
we can say $D \approx 0.6 L$,
while $D \approx 0.8 L$ when
$L = 64$ according to Fig.~\ref{Fig:Pplt} (b).
Furthermore, $D$ also depends on the SNR.
In addition, as mentioned earlier, although $K_m > D$,
it is also possible that some signals can be recovered
as shown in Fig.~\ref{Fig:Pplt}.
If a better recovery algorithm is used, we expect to have
a larger $D$, which results in a better performance.
In \cite{Choi18}, the performances of different algorithms can be found,
while we only consider S-OMP (for simulations) in this paper.

In Fig.~\ref{Fig:Pplt}, we also show the number of unsuccessful
devices of 
conventional multichannel ALOHA, which is given by
$U_{\rm aloha} (K_m) = K_m - K_m \left( 1 - \frac{1}{L} \right)^{K_m-1}$,
where each RB has only $L$ (not $N$) orthogonal\footnote{In order to
increase the number of channels for a high
throughput in multichannel ALOHA,
non-orthogonal SCs can be used. In this case,
a bank of correlators cannot be used at a receiver due to 
interference and a CS based MUD approach needs to be
used as in \cite{Guo08, Zhu11}, which results in CRA. Thus,
by conventional multichannel ALOHA, we mean multichannel ALOHA
with $L$ orthogonal channels in this paper.}  SCs.
In conventional multichannel ALOHA, it is not
necessary to use CS based MUD due to orthogonal SCs. That is,
a bank of correlators can be used to recover multiple signals.
Although there is no interference, due to a smaller number of channels
(not $N$, but $L$),
the probability of collision is higher than that of CRA.
As a result, as shown in 
Fig.~\ref{Fig:Pplt}, the performance of conventional multichannel ALOHA
would be worse that that of CRA.
From this, we can claim that the advantage of CRA
over conventional multichannel ALOHA is a low probability
of collision thanks to an increased number of  (non-orthogonal virtual)
channels (from $L$ to $N$) at the cost of high recovery
complexity at a receiver.
Note that as we only consider S-OMP in this paper,
it might be useful to consider the computational complexity
of S-OMP. Since S-OMP is a greedy algorithm \cite{Tropp06_SP} \cite{Eldar12},
its computational complexity is not significantly high, which is
mainly $O ( c L^2 + N)$ per RB,
where $c > 1$ is a constant.
On the other hand, the complexity per RB at a receiver in conventional
multichannel ALOHA might be $O(L^2)$.

It is important to note that the receiver complexity of MRB-CRA 
with S-OMP decreases with $M$ when $J$ is fixed.
Since the total computational complexity is
$O(M (c L^2 + \eta L)) = O \left( c \frac{J^2}{M} + \eta J\right)$,
we can clearly see that a large $M$ (or more  RBs)
is desirable for a low computational
complexity.
However, as shown in Fig.~\ref{Fig:Pplt}, if $L$ decreases,
$\frac{D}{L}$ tends to decrease, which degrades
the recovery performance. Thus, we can see that there might be a
trade-off between the performance and complexity
(this will be confirmed by simulations in Section~\ref{S:Sim}).

\section{Stability Analysis with Fast Retrial}	\label{S:SA}

In \cite{YJChoi06}, for multichannel ALOHA,
a re-transmission strategy which is called fast retrial
is proposed, where a collided packet is re-transmitted at the next
slot without any random backoff delay thanks
to the existence of multiple channels.
Since we assume multiple RBs for CRA, i.e., MRB-CRA, 
fast retrial can be employed
as a re-transmission strategy for collided SCs as mentioned in
Section~\ref{S:SM}.
In this section, we study the stability of MRB-CRA with fast retrial.

Suppose that a device experiencing SC collision
attempts re-transmission in the next
slot based on fast retrial \cite{YJChoi06}. 
For re-transmission,
the device chooses
an RB among $M$ RBs and an SC within the selected
RB uniformly at random.
There are also new active devices.
For convenience, let $A_m (q)$ denote the number of new active devices
to RB $m$ at time slot $q$. Here, $q$ is used for 
the time
slot index.
Thus, $K_m (q)$ is a sum of 
the numbers of new packets\footnote{We assume that
packets and active devices are interchangeable in this section
as an active device transmits a packet.}
and re-transmitted packets,
where $K_m (q)$ represents the number of active devices
transmitting packets to RB $m$ at time slot $q$.
Throughout the paper,
we consider 
an independent Poisson random variable for $A_m(q)$ as follows:
\be
\Pr(A_m(q) = n) = \frac{\lambda^n e^{-\lambda}}{n!},
	\label{EQ:A_P}
\ee
where $\lambda$ is the arrival rate per RB,
i.e., $A_m (q) \sim {\rm Pois} (\lambda)$. Here,
${\rm Pois} (\lambda)$ denotes the distribution
of a Poisson random variable with mean $\lambda$.

In fast retrial, thanks to multiple RBs, we consider the following 
rate control strategy.
\begin{itemize}
\item[{\bf A3}] If $K_m(q) > \bar K$, the AP informs the devices not to 
send any new packets, where $\bar K \ (> D)$
is a pre-determined threshold.
\end{itemize}
This rate control strategy is necessary to avoid
the growth of $K_m (q)$ so that the access delay cannot be arbitrarily long.
Note that there can be other rate control strategies. However,
for simplicity, we only consider {\bf A3}.
According to {\bf A3}, 
the AP needs to broadcast the binary signals that inform
the states of RBs (whether or not $K_m(q)$ is greater than $\bar K$).
Thus, at the beginning of
each (uplink) time slot, there should be a broadcast signal 
of $M$ bits from the AP to devices
for the rate control.

Let $\bk (q) = [K_1 (q) \ \ldots \ K_M (q)]^\rT$.
Under the assumptions of {\bf A1} and {\bf A2},
we define the number of unsuccessfully recovered packets
in RB $m$ at time slot $t$ as 
\be
[K_m (q)]_D = \left\{
\begin{array}{ll}
K_m (q), & \mbox{if $K_m (q) > D$;} \cr
i, & \mbox{if $K_m (q) \le D$ w.p. 
$\beta(K_m(q), n)$,}
\cr
\end{array}
\right.
\ee
where
$\beta (K_m,i)$ represents the probability
that there are $i$
collided SCs when there are $K_m \ (\le N)$ packets 
and each packet independently chooses one among $N$ SCs.
Clearly, $[K_m (q)]_D$ is a random variable.
%In general, $\beta(K_m,n)$ is not easy to find unless $K_m$ is sufficiently
%small. 
%For example, if $K_m = 1$, we have $\beta(1,0) = 1$ and $\beta(1,1) = 0$.
%If $K_m = 2$, it can be shown that
%\begin{align*}
%\beta(2,0) & = 1 - \frac{1}{N} \cr
%\beta(2,1) & = 0 \cr
%\beta(2,2) & = \frac{1}{N}.
%\end{align*}

With fast retrial, the unsuccessfully recovered packets are 
scheduled to re-transmit in the next time slot.
The number of re-transmitted packets 
from RB $m$ at time slot $q$ to RB $l$ at time slot $q+1$ 
is denoted by $\cR_{m,l} (i)$, where $i$ represents the 
number of unsuccessfully recovered packets in RB $m$ at time slot $q$.
Then, the total number of packets 
in RB $m$ at time slot $q+1$ can be expressed as
\be
K_m (q+1) = \sum_{l = 1}^M \cR_{m,l} ([K_l (q)]_D) + 
\tilde A_m (q+1),
	\label{EQ:KRK}
\ee
where 
\be
\tilde A_m (q+1) = \left\{
\begin{array}{ll}
A_m (q+1), & \mbox{if $K_m(q) \le \bar K$};\cr
0, & \mbox{o.w.}\cr
\end{array}
	\label{EQ:tA}
\right.
\ee
Note that \eqref{EQ:tA} is due to the rate control of {\bf A3}.
From \eqref{EQ:KRK} and \eqref{EQ:tA}, we can see that $\bk(q)$ becomes
a Markov process.
We derive a sufficient condition for stability
as follows.

\begin{mytheorem}	\label{T:1}
Let $\bar A_m = \uE[A_m (q)]$. With the rate control of {\bf A3},
if 
\be
\bar A_m = \lambda < 
B_{D,N} \deft D \left(1 - \frac{1}{N} \right)^{D-1}, \ \mbox{for all $m$},
	\label{EQ:AD}
\ee
then $\bk(q)$ is positive recurrent.
\end{mytheorem}
\begin{IEEEproof}
See Appendix~\ref{A:T1}.
\end{IEEEproof}

In \eqref{EQ:AD}, it is noteworthy that
$B_{D, N} \approx D$ if $N \gg  D$.
This indicates that the new arrival rate
per RB is to be less than $D$ for stable MRB-CRA when $N \gg L > D$.
This indicates that the estimation of $D$
plays a crucial role in MRB-CRA with fast retrial.

\section{Performance Analysis of MRB-CRA}	\label{S:PA}

In this section, 
in order to find the throughput and delay of MRB-CRA,
we consider a steady state analysis 
with a receiver that is capable of recovering $D$ multiple signals.
%In addition, for tractable analysis, we consider approximations
%with underloaded traffics.

\subsection{A Steady State Analysis}

In general, the throughput analysis of 
MRB-CRA with fast retrial is not easy due to the interaction
between multiple RBs in terms of their numbers of
the packets to be re-transmitted.
That is, $K_m(q+1)$ depends on
$K_l (q)$, $l = 1,\ldots, M$, as shown in \eqref{EQ:KRK}.
To avoid this difficulty, in the steady state, we assume that
the number of packets to be re-transmitted 
(or collided packets in the previous time slot) in each RB 
is an independent Poisson random variables with mean $\lambda_2$,
i.e., 
$$
\sum_{l=1}^M \cR_{m,l} ( [K_l (q)]_D ) 
\sim {\rm Pois} (\lambda_2), \ m = 1,\ldots,M,
$$
where $\lambda_2  =
\uE\left[
\sum_{l=1}^M \cR_{m,l} ( [K_l (q)]_D ) \right]$.
Clearly, this assumption is not true. However,
if $M$ is sufficiently large,
it might be a reasonable approximation such as
the Kleinrock independence approximation \cite{Kleinrock64} 
\cite{BertsekasBook}.

\begin{figure}[thb]
\begin{center}
\includegraphics[width=\figwidth]{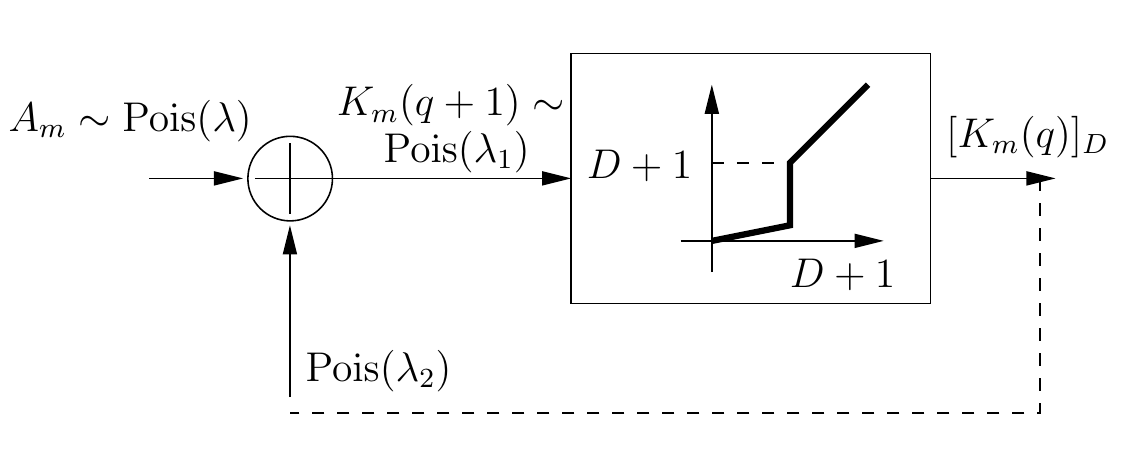}
\end{center}
\caption{An illustration
for the relation between the average number of packets to be transmitted
and that of collided packets per RB, which are denoted by
$\lambda_1$ and $\lambda_2$, respectively.}
        \label{Fig:Fig1}
\end{figure}

Suppose that the arrival rate
is sufficiently low so that $\tilde A_m (q) = A_m (q)$ for all $q$,
i.e., the rate control of {\bf A3} is hardly imposed.
Although this is an approximation, 
we assume this for tractable analysis (i.e., the approximation
allows us to simplify the analysis, while simulation
results in Section~\ref{S:Sim} show that this approximation
is reasonable).
Since the packets to be re-transmitted
are to be re-allocated to the RBs uniformly at random,
from \eqref{EQ:KRK},
we assume that 
$K_m (q+1)$ becomes a Poisson 
random variable\footnote{Since the sum of two independent
Poisson random variables with means
$\lambda_a$ and $\lambda_b$ is a Poisson random variable
with the mean $\lambda_a + \lambda_b$, we can assume
that $K_m (q+1)$ is also a Poisson random variable.}
with mean $\lambda + \lambda_2$.
For convenience, let
\be
\lambda_1 = \lambda + \lambda_2,
	\label{EQ:T1}
\ee
which is the average number of packets to be transmitted per RB
as illustrated in Fig.~\ref{Fig:Fig1}.

\begin{mytheorem}	\label{T:X}
Suppose that 
$K_m (q) \sim {\rm Pois} (\lambda_1)$.
Under {\bf A1} and {\bf A2}, 
$\lambda_2$ can be given by
\begin{align}
\lambda_2 
= \lambda_1 \left( 1 - e^{-\frac{\lambda_1}{N}}
\frac{\Gamma(D,\nu_1)}{(D-1)!} \right),
	\label{EQ:T2}
\end{align}
where $\nu_1 = \lambda_1 \left( 1  - \frac{1}{N} \right)$
and $\Gamma(d, x) = \int_x^\infty
t^{d-1} e^{-t} dt$ is the upper incomplete gamma function.
\end{mytheorem}
\begin{IEEEproof}
See Appendix~\ref{A:X}.
\end{IEEEproof}

To satisfy \eqref{EQ:T1} and \eqref{EQ:T2},
we can show that
$\lambda$ has to be bounded as
\begin{align}
\lambda 
\le \lambda_{\rm max} 
= \max_{\lambda_1} 
\lambda_1 
 e^{-\frac{\lambda_1}{N}}
\frac{\Gamma(D, \nu_1)}{(D-1)!},
	\label{EQ:lmax}
\end{align}
where $\lambda_{\rm max}$ denotes the maximum of $\lambda$.
In addition, denote by $\lambda_1^*$ the solution to \eqref{EQ:lmax}.
Clearly, if $\lambda$ is less than or equal to 
$\lambda_{\rm max}$, the steady state solution that satisfies
\eqref{EQ:T1} and \eqref{EQ:T2} exists.

Since $e^{-\lambda_1/N} < 1$,
from \eqref{EQ:lmax},
an upper-bound on $\lambda_{\rm max}$ can be found as follows:
\be
\bar \lambda_{\rm max} 
= \max_{\lambda_1} \lambda_1 
\frac{\Gamma(D, \nu_1)}{(D-1)!}  \ge \lambda_{\rm max}.
	\label{EQ:llmax}
\ee
Note that as $N$ increases, 
$\bar \lambda_{\rm max}$ becomes tighter and
approaches $\lambda_{\rm max}$.

\begin{mytheorem}	\label{T:2}
For $D > 1$,
there exists $N > 0$ such that
\be
\bar \lambda_{\rm max} < B_{D,N},
	\label{EQ:lBDN}
\ee
which guarantees stable MRB-CRA with fast
retrial (according to Theorem~\ref{T:1}) for
$\lambda \le \lambda_{\rm max}$
(since $\lambda_{\rm max} < \bar \lambda_{\rm max}$).
\end{mytheorem}
\begin{IEEEproof}
See Appendix~\ref{A:T2}.
\end{IEEEproof}

Consequently, for a given $D$,
we can find the maximum arrival rate,
$\lambda_{\rm max}$ from \eqref{EQ:lmax}.
As long as the arrival rate is lower than or equal to
$\lambda_{\rm max}$, we can also guarantee stable MRB-CRA with fast
retrial according to Theorems~\ref{T:1} and ~\ref{T:2}.
In fact, with $\lambda \le \lambda_{\rm max} \
(< \bar \lambda_{\rm max} < B_{D,N})$,
we can also expect that the rate control of {\bf A3} may not be frequently
used as $K_m (q)$ might stay around $\lambda_1$ which is smaller
than $D$, i.e., most transmitted packets become successful packets.
However, to decide $\lambda$,
we first estimate $D$ based on the performance of CS based MUD
for a given set of parameters (e.g., $L$, $N$, and SNR).
Thus, as mentioned earlier, the estimation of $D$ 
is important in MRB-CRA.

\subsection{Throughput and Delay in the Steady State}

In the steady state, 
for a given $\lambda$ satisfying \eqref{EQ:lmax},
we can find $\lambda_1$
by solving \eqref{EQ:T1} and \eqref{EQ:T2} as shown above.
With $\lambda_1$,
under {\bf A1} and {\bf A2},
the throughput per RB
can be given by
\begin{align}
T_{\rm cra} & = \sum_{n=0}^D \frac{e^{-\lambda_1} \lambda_1^n}{n!}
n \left(1 - \frac{1}{N} \right)^{n-1} \cr
& = \lambda_1 e^{-\lambda_1} \sum_{n=0}^D \frac{1}{n!}
\left( \lambda_1 \left(1 - \frac{1}{N} \right) \right)^n \cr
& \le \lambda_1 e^{-\frac{\lambda_1}{N}} ,
\end{align}
where the upper-bound becomes tight when $D - \lambda_1$ is large.
Recall that $\eta = N/L$. Then, it follows
\begin{align}
T_{\rm cra} \le \lambda_1 e^{-\frac{\lambda_1}{N}} 
= \lambda_1 e^{- \frac{\Lambda_1}{\eta J} },
	\label{EQ:TMT}
\end{align}
where $\Lambda_1 = M \lambda_1$.

We can find the average access
delay from the relationship between $\lambda_1$
and $\lambda$ as shown in Fig.~\ref{Fig:Fig1}.
The average number of packets to be transmitted
(in a time slot), $\lambda_1$,
can be seen as the number of devices in a system, while
$\lambda$ is the average number of devices (within a time slot)
entering into the system.
Based on Little's law \cite{HajekBook},
the average access delay (in the number of slots) can be found 
as
\be
\tau = \frac{\lambda_1}{\lambda}.
	\label{EQ:tau}
\ee

\subsection{Comparison with Multichannel ALOHA}

It might be interesting to compare the throughput of MRB-CRA
with that of conventional multichannel ALOHA.

Suppose that there are $L$ orthogonal
channels 
per RB in conventional
multichannel ALOHA.
In an RB, when there are $n$ active devices,
the number of successfully transmitted packets is
given by $n \left(1 - \frac{1}{L} \right)^{n-1}$.
Thus, the average number of successfully transmitted packets is
\begin{align}
T_{\rm aloha} 
= \sum_{n} n \left(1 - \frac{1}{L} \right)^{n-1}
\frac{\lambda^n e^{-\lambda}}{n!} 
= \lambda e^{-\frac{\lambda}{L}}.
\end{align}
According to \cite{Shen03}, the maximum stable throughput
(per RB)
becomes 
\be
T_{\rm aloha}
\le \hat T_{\rm aloha} = L e^{-1}.
\ee

Let $\Lambda = M \lambda$, which is the total arrival rate
to the system. In MRB-CRA, since $D < L$, 
from \eqref{EQ:lmax} and \eqref{EQ:lBDN},
we can have the following inequality:
$$
\lambda = \frac{\Lambda}{M} < B_{D, N} < L,
$$
which shows that the total arrival rate, $\Lambda$, cannot be 
greater than $J$, i.e.,
\be
\Lambda < M L = J.
	\label{EQ:LMJ}
\ee
However, if $\Lambda$ is sufficiently close to $J$ so that
$\Lambda_1 = J$ 
(note that since $\Lambda < \Lambda_1$, the inequality
in \eqref{EQ:LMJ} is valid in this case),
from \eqref{EQ:TMT}, we have
\be
T_{\rm cra} \approx \lambda_1 e^{-1/\eta} \le L e^{-1/\eta}.
\ee

As $\eta \gg 1$, we have $e^{-1/\eta} > e^{-1}$, from which 
it can be claimed that
the throughput of MRB-CRA can be higher than that of conventional multichannel
ALOHA. 
For example, under
the optimistic assumption that $\lambda_1$ is sufficiently close to $L$,
if $\eta \ge 3.258$, we can claim that the throughput of MRB-CRA
can be higher than that of multichannel ALOHA by
at least a factor of 2.

It is noteworthy that in \cite{Choi17_Stability},
it is shown that the throughput of MRB-CRA can be about two-time higher
than that of multichannel ALOHA when a certain controlled access
probability strategy is employed.
Unfortunately, in this paper, we cannot find the throughput of MRB-CRA
with fast retrial as a closed-form expression (i.e.,
no closed-form expression for $\lambda$ is obtained).
Thus, we can consider 
simulations 
(as shown in Fig.~\ref{Fig:Pplt} 
and in the next section)
for comparisons with conventional
multichannel ALOHA.

\section{Simulation Results}	\label{S:Sim}

In this section, we present simulation results
for MRB-CRA with fast retrial.
For simulations, we consider
$\bC$ whose elements are
independent CSCG random variables with zero mean and variance
$\frac{1}{L}$. As mentioned earlier, the S-OMP 
algorithm is used for CS based MUD to detect multiple signals at an AP.
To see the performance, we mainly
consider the average number of transmitted packets per RB,
$\uE[K_m (q)]$, and
the average number of successful packets per RB,
which might be identical to $\lambda$
(wile $\lambda_1 =\uE[K_m (q)]$) in the steady state for 
stable MRB-CRA.
From \eqref{EQ:tau}, for short access delay,
we expect that 
the average number of transmitted packets per RB
is not too larger than that of successful packets per RB.

\begin{figure}[thb]
\begin{center}
\includegraphics[width=\figwidth]{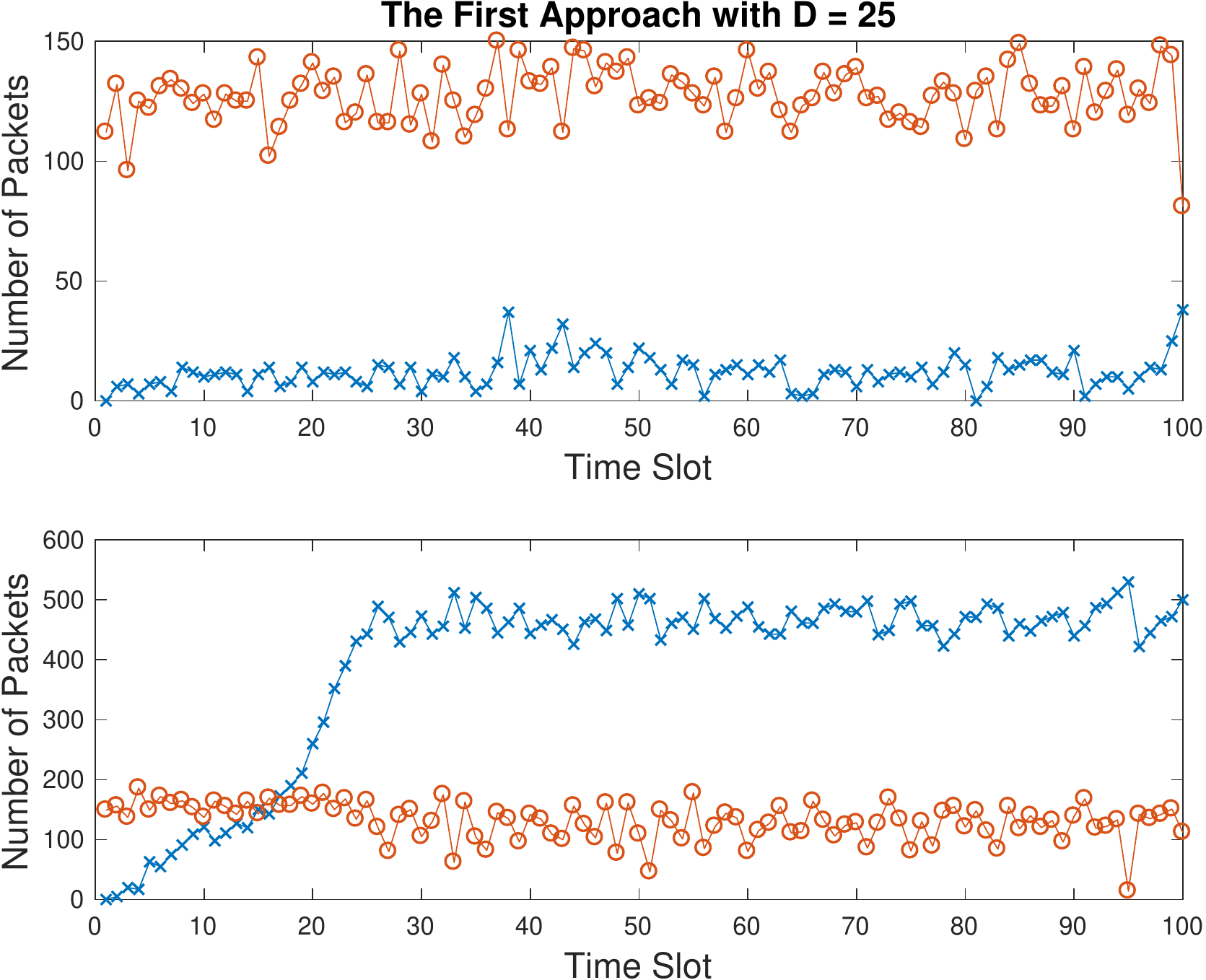}
\end{center}
\caption{Evolutions of 
$\sum_m K_m (q)$
and $\sum_m A_m (q)$ over time slots
when $(L, M) = (32, 8)$, $\eta = 10$, $\bar K = 2L$, and SNR $= 20$ dB.
The curves with cross ($\times$) marks are $\sum_m K_m (q)$
and those with circle ($\circ$) marks are $\sum_m A_m (q)$.}
        \label{Fig:evol}
\end{figure}

We consider two different approaches to decide the arrival
rate per RB, $\lambda$. In the first approach,
we fix $D$ for given $(L, N)$ and decide $\lambda$ 
to be $\lambda_{\rm max}$ in \eqref{EQ:lmax}.
In the second approach, $\lambda$ is directly decided.
In Fig.~\ref{Fig:evol},
we present the evolution of $\sum_{m} K_m (q)$
and $\sum_m A_m (q)$ over time slots when 
$(L, M) = (32, 8)$, $\eta = 10$, $\bar K = 2L$, and SNR $= 20$ dB.
In the upper figure in 
Fig.~\ref{Fig:evol}, the first approach is used to decide
$\lambda$ with $D = 25$. In this case, $\lambda$ becomes
$16.04$. 
On the other hand, 
in the lower figure in
Fig.~\ref{Fig:evol}, the second approach is used with $\lambda = 20$.
We can see that the determination of $D$ can help decide
a proper arrival rate, $\lambda$, to keep $K_m (q)$ low
so that the access delay is short.
On the other hand, if $\lambda$ happens to be high,
it may result in large $K_m (q)$'s and unstable MRB-CRA. 
Note that $\lambda = 20$
corresponds to about $D = 31$ from \eqref{EQ:lmax}.
Thus, in order to support $\lambda = 20$, CS based MUD or
the recovery algorithm
should be capable of detecting $D = 31$ multiple signals.
If we use an ideal method, it might be achievable
according to \eqref{EQ:tL1} as $L = 32$.
However, the S-OMP algorithm is a greedy algorithm that has a suboptimal
performance. Thus, a more realistic value for $D$ might 
be considered. From Fig.~\ref{Fig:evol}, $D = 25$ seems a reasonable
estimate of $D$ to decide $\lambda$.

\begin{figure}[thb]
\begin{center}
\includegraphics[width=\figwidth]{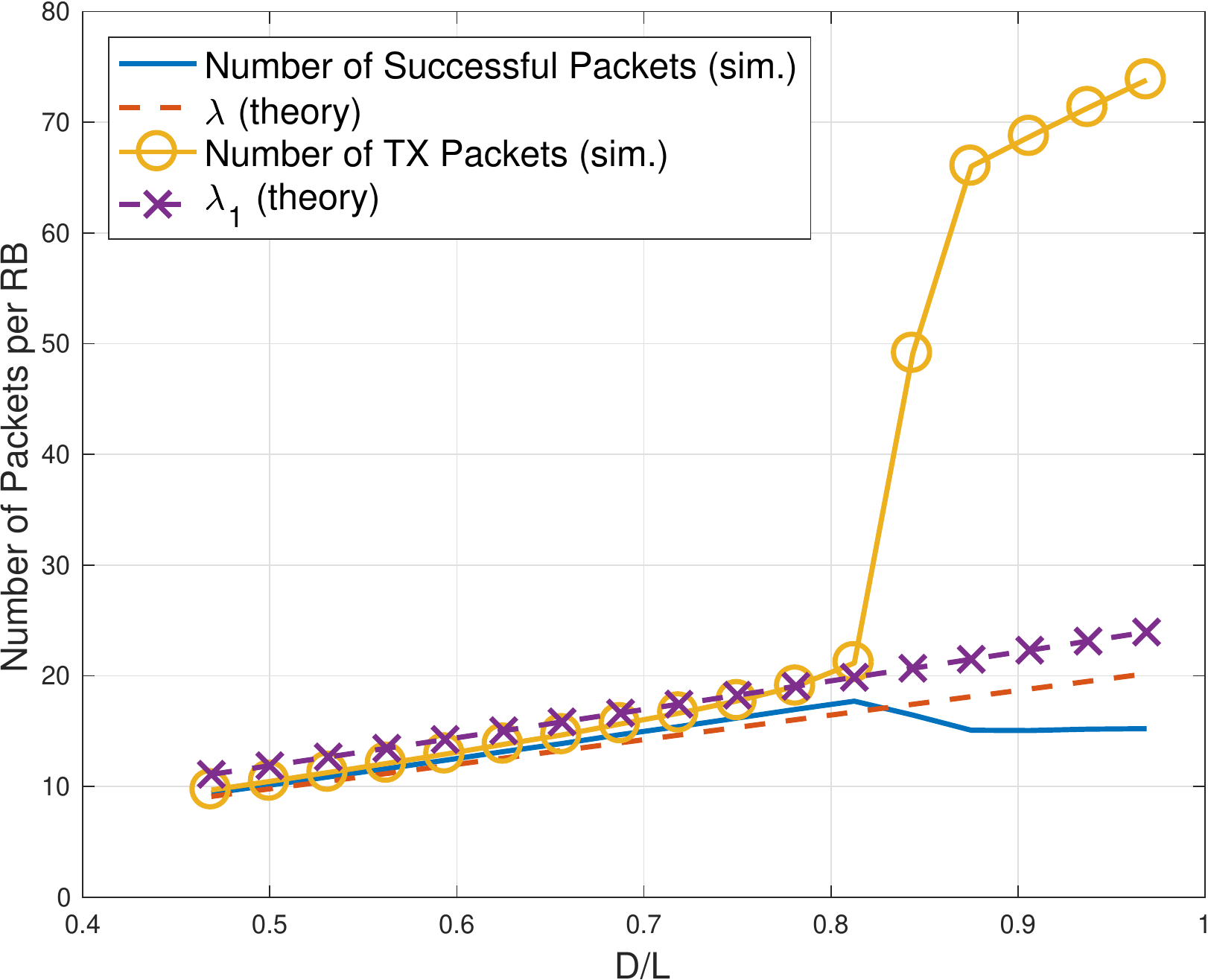} \\
(a) \\
\includegraphics[width=\figwidth]{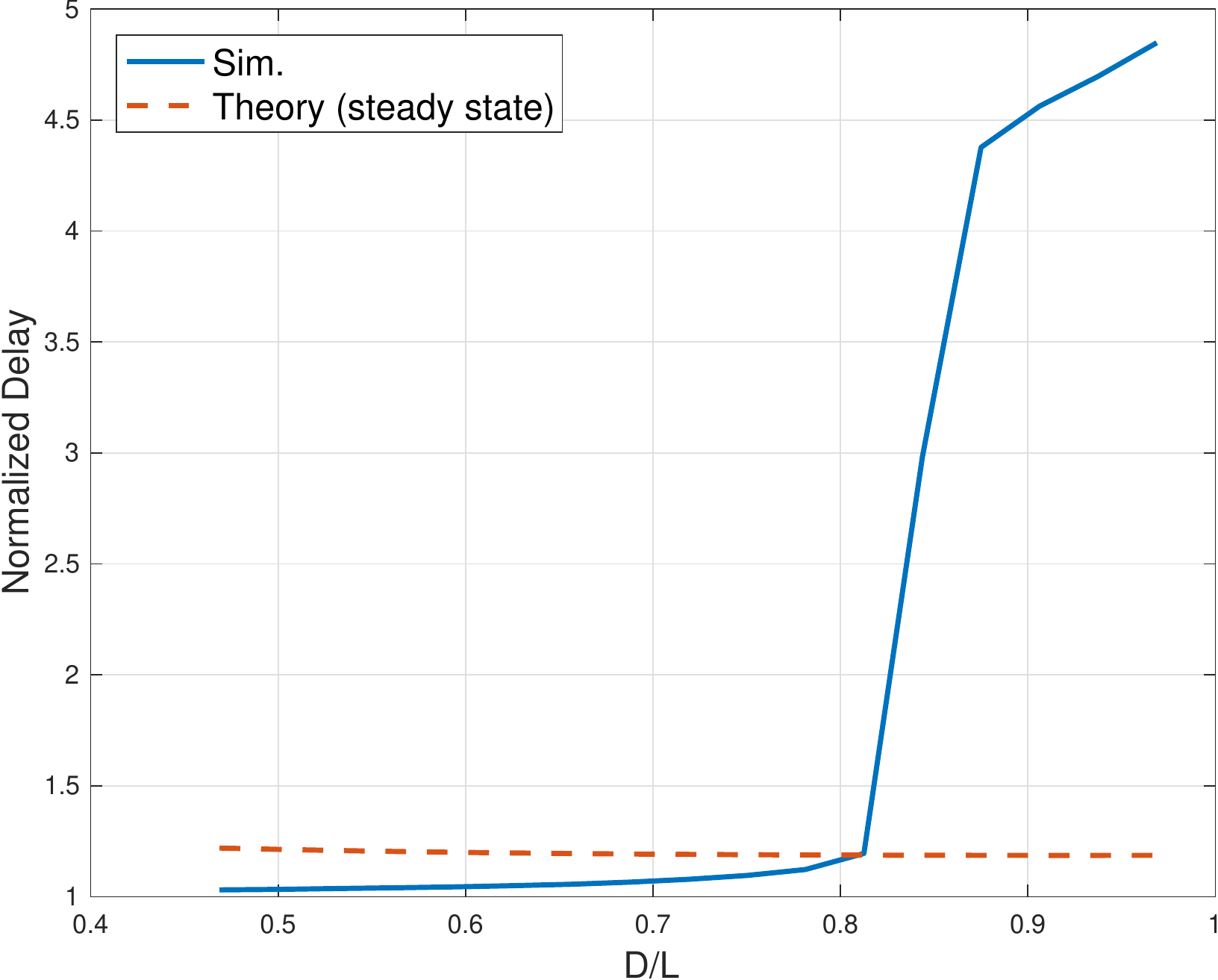} \\
(b) 
\end{center}
\caption{Performance of MRB-CRA with fast retrial for 
different values of $D$ to decide $\lambda$
when $(L, M) = (32, 8)$, $\eta = 10$, $\bar K = 2D$, and SNR $= 20$ dB.
(a) the number of successful packets
(or successfully recovered packets) 
and the number of transmitted packets per RB;
(b) normalized access delay.}
        \label{Fig:plt1}
\end{figure}

Fig.~\ref{Fig:plt1} (a) shows 
the number of successful packets
(or successfully recovered packets) 
and the number of transmitted packets per RB
for different values of $D$ to decide $\lambda$
according to the first approach
(i.e., using \eqref{EQ:lmax} for given $D$)
when $(L, M) = (32, 8)$, $\eta = 10$, $\bar K = 2D$, and SNR $= 20$ dB.
This shows that if the value of $D$ is less than
80\% of $L$ (i.e., $D \le 26$), 
the arrival rate might be sufficiently low so that
$K_m (q)$ does not significantly grow.
In other words, $D \le 26$ might be an underestimate of the recovery
performance.
On the other hand, if $D$ is decided to be too high or
the performance of CS based MUD is overestimated,
$K_m (q)$ grows 
and the rate control of {\bf A3} has to be imposed to avoid
any excessive delay.
In Fig.~\ref{Fig:plt1} (b), we present the normalized
access delay, which is the ratio of 
the number of transmitted packets to
the number of successful packets as in \eqref{EQ:tau},
where it is shown that
an overestimate of $D$ results in a relatively long access delay.

\begin{figure}[thb]
\begin{center}
\includegraphics[width=\figwidth]{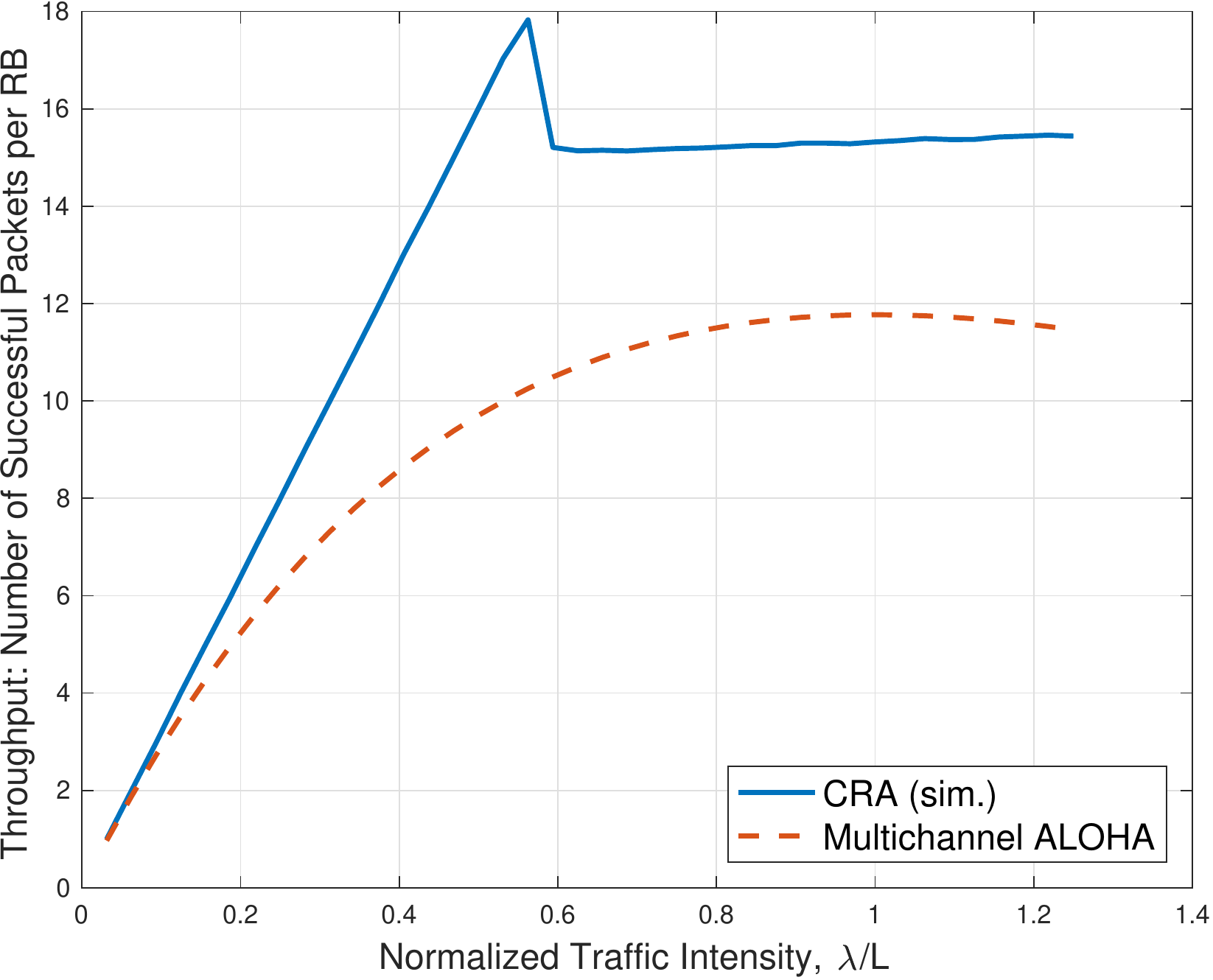} \\
(a) \\
\includegraphics[width=\figwidth]{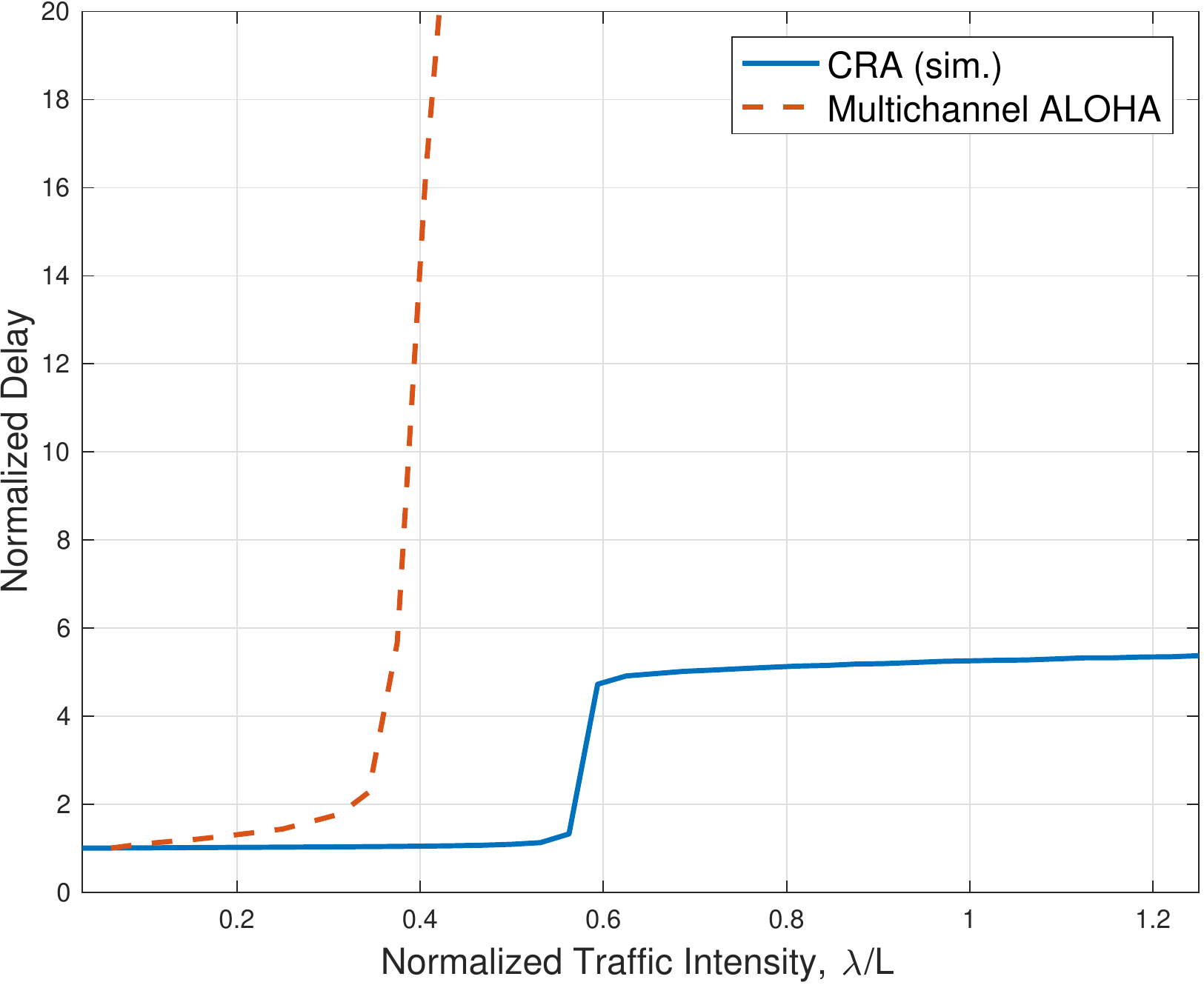} \\
(b) \\
\end{center}
\caption{Performances of MRB-CRA and conventional ALOHA
for different values of arrival rate, $\lambda$
when $(L, M) = (32,8)$, $\eta = 10$, $\bar K = 2L$, and SNR $=20$ dB:
(a) throughput (or the average number of successful packets per RB);
(b) normalized access delay.}
        \label{Fig:plt3}
\end{figure}

For comparison with 
conventional multichannel ALOHA,
the throughput (i.e., the average number of successful packets per RB)
is shown for different values of arrival rate,
$\lambda$,
in Fig.~\ref{Fig:plt3}
when $(L, M) = (32,8)$, $\eta = 10$, $\bar K = 2L$, and SNR $=20$ dB.
For conventional multichannel ALOHA, we consider
the controlled access probability 
proposed in \cite{Galinina13} 
with the assumption that the number of transmitted packets
at each slot is available at the devices to decide their access probability.
We can see that MRB-CRA outperforms conventional multichannel ALOHA
in terms of both the throughput and access delay.

Note that in Fig.~\ref{Fig:plt3}, the throughput and access delay
of MRB-CRA behave differently when $\lambda/L \ge 0.6$ or $\lambda \ge 19$. 
As mentioned earlier, if $\lambda$ is too high,
the rate control of {\bf A3} is imposed to keep the access delay
reasonable, which, however, results in a lower throughput.

\begin{figure}[thb]
\begin{center}
\includegraphics[width=\figwidth]{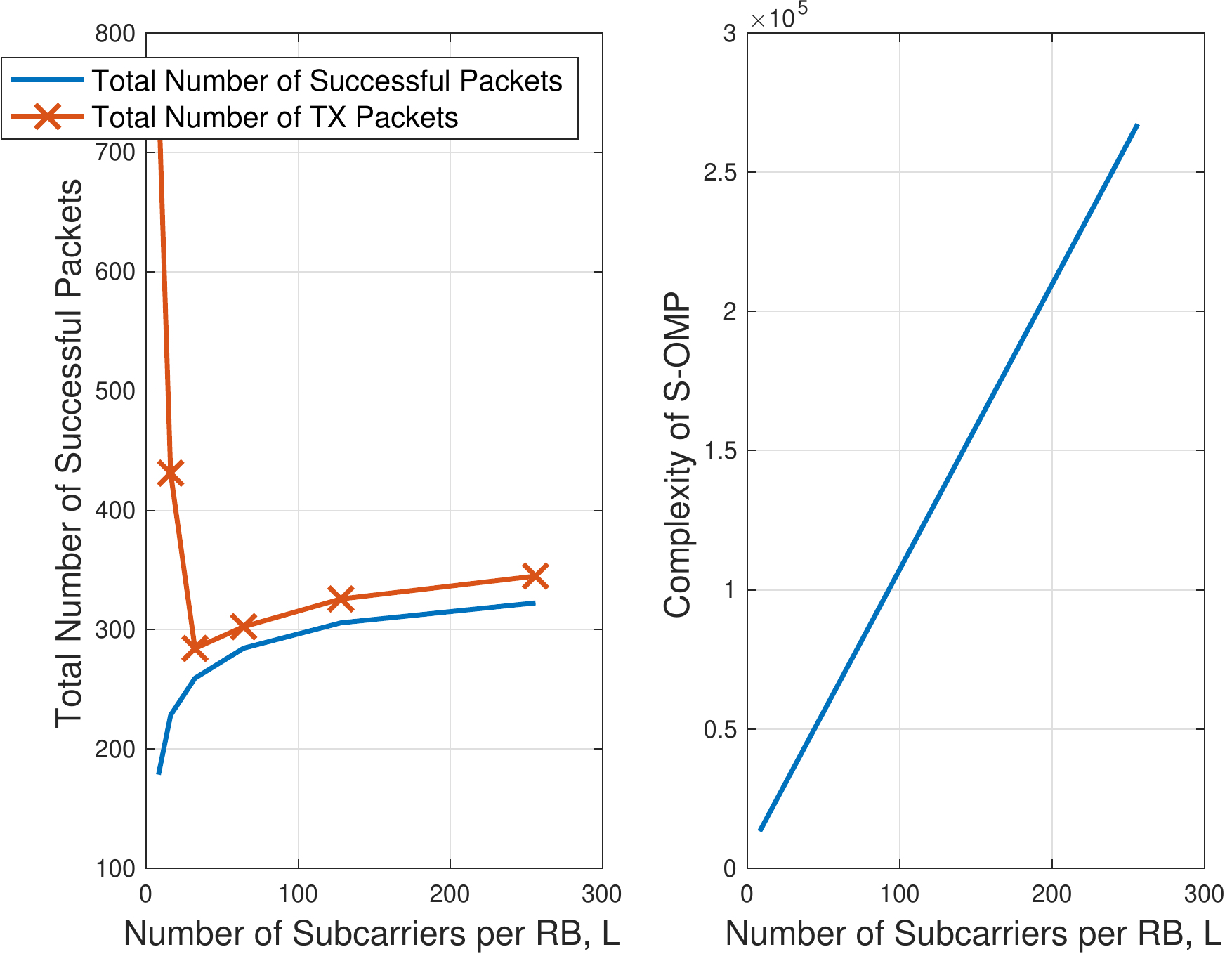} \\
\hskip 0.5cm (a) \hskip 3.5cm (b) 
\end{center}
\caption{Performances of MRB-CRA
with different numbers of $L$ for a fixed $J$ when
$J = 512$, SNR $=20$ dB and $\Lambda = 0.8J$:
(a) the total number of successful packets
and the total number of transmitted packets;
(b) the total complexity.}
        \label{Fig:plt4}
\end{figure}

In order to see the impact of $L$ on the performance,
we show the total number of successful
packets and the total number of transmitted packets (i.e.,
$\uE \left[\sum_m K_m (q)\right]$) as well as the complexity
for different values of $L$ 
when the total number of
subcarriers is fixed as $J = 512$ in Fig.~\ref{Fig:plt4} 
with SNR $=20$ dB and $\Lambda = 0.8J$
(or $\lambda = 0.8 L$).
As mentioned earlier, the performance
of S-OMP is improved as $L$ increases, which means that $D$ can increase
with $L$ (which can be confirmed by Fig.~\ref{Fig:Pplt}).
Thus, for given $\lambda = 0.8 L$, 
$D$ should be at least greater than $0.8 L$
for reasonable performances of MRB-CRA 
or stable MRB-CRA (with reasonably short access delay). 
As shown in Fig.~\ref{Fig:plt4} (a), we need $L \ge 32$
for reasonably good performances. 
Furthermore, as $L$ increases, we can see that the performance
is improved as expected. However, 
Fig.~\ref{Fig:plt4} (b) shows that the complexity
increases with $L$ (as $M$ decreases with $L$ since $J$ is fixed).
Thus, one big RB is not desirable for CRA due to high computational
complexity.
As mentioned earlier, this exhibits the trade-off between
the performance and complexity.

\begin{figure}[thb]
\begin{center}
\includegraphics[width=\figwidth]{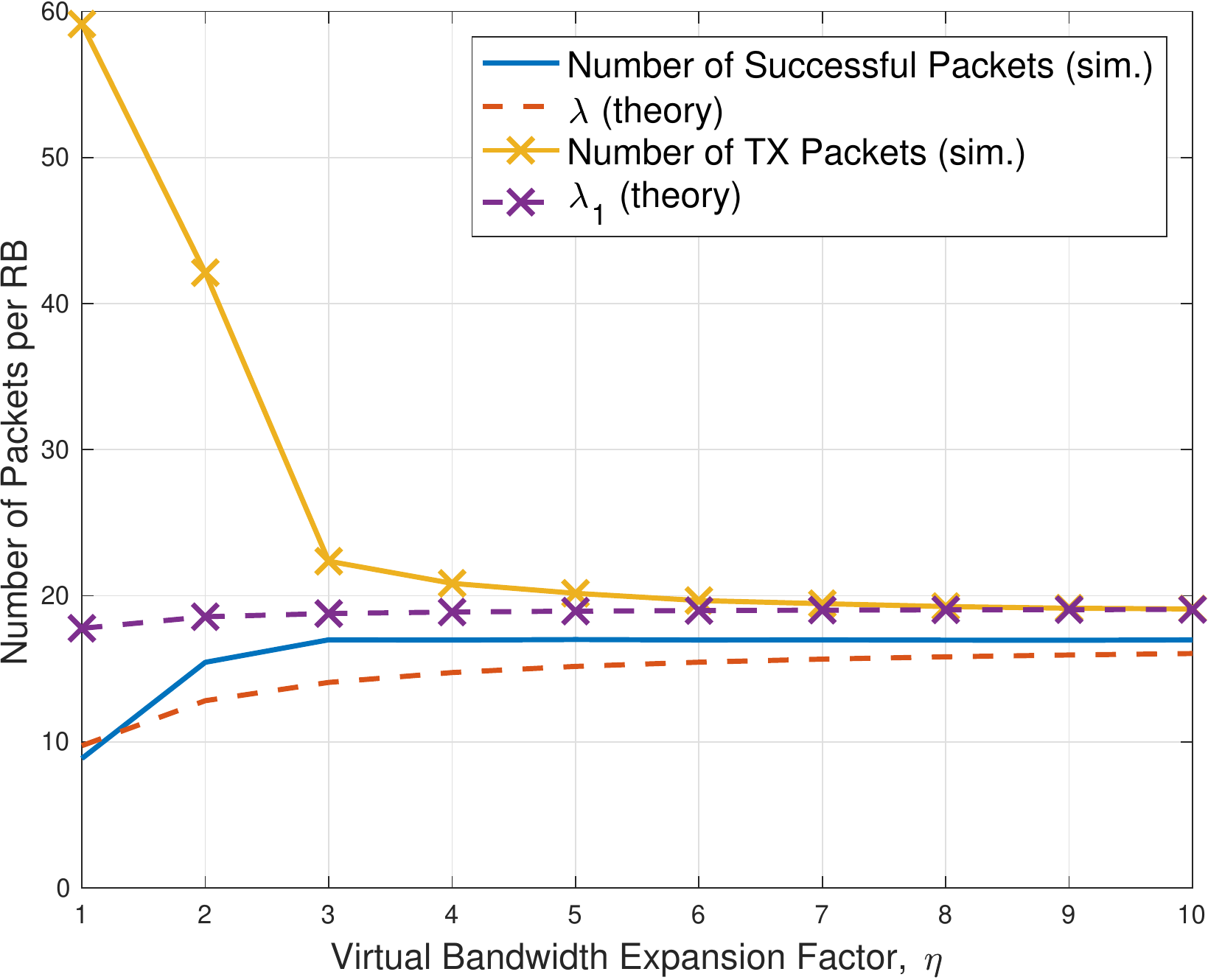}
\end{center}
\caption{The number of successful packets
and the number of transmitted packets in the steady state
for different values of $\eta$
when $(L, M) = (32,8)$, $D = 25$, and SNR $=20$ dB.}
        \label{Fig:plt6}
\end{figure}

Fig.~\ref{Fig:plt6}, we show
the number of successful packets
and the number of transmitted packets
for different values of the virtual bandwidth expansion
factor, $\eta$,
when $(L, M) = (32,8)$, $D = 25$, and SNR $=20$ dB.
It is shown that as $\eta$ increases, the number of
transmitted packets per RB, $\uE[K_m(q)]$, decreases. Thus,
it is desirable to have a large
$\eta$ (i.e., greater than 5)
to generate more virtual channels per RB to reduce the probability of
collision and result in a better performance.
However, since $N$ increases with $\eta$, the increase of $\eta$
results in  a higher computational complexity. Thus,
$\eta$ should not be too large. From
Fig.~\ref{Fig:plt6}, $\eta = 6$ or 7 seems a reasonable
choice to provide a good performance with a relatively low complexity.

\section{Concluding Remarks}	\label{S:Conc}

We have proposed an MRB-CRA scheme with fast
retrial in this paper. It was shown that
the proposed scheme is not only scalable, but also
computationally efficient in terms of receiver's complexity
due to multiple RBs. We carried out
the stability analysis based on
the Foster-Lyapunov stability criterion,
and derived a maximum arrival rate 
for stable MRB-CRA with fast retrial. Furthermore,
we studied a steady state analysis
to find the throughput and delay of MRB-CRA.
Through analysis and simulation results,
it was shown that the proposed MRB-CRA scheme can have a higher
throughput than
conventional multichannel ALOHA with a lower
access delay and enjoy a trade-off between
the performance and complexity in terms of the number of RBs.

\appendices

\section{Proof of Theorem~\ref{T:1}}	\label{A:T1}

To show that $\bk (q)$ is positive recurrent,
we can use the notion of Lyapunov function that is a nonnegative
potential function \cite{HajekBook}.
In particular, from a Lyapunov function, we can have the drift function,
which represent the variation of Lyapunov function in time.
If the drift function satisfies certain conditions,
we can show that $\bk (q)$ is positive recurrent.

Let $V(q) = \sum_{m=1}^M K_m (q)$ be the Lyapunov function,
which is a nonnegative function for 
$$
\Omega = \{\bk = [K_1 \ \ldots \ K_M]^\rT
\,|\, K_m \in \{0, \ldots, M\bar K\} \}.
$$
Note that $\Omega$ is a finite set 
as the maximum 
value of $K_m (q)$ is $M \bar K$ due to the assumption of {\bf A3}.
Based on Foster's theorem \cite{HajekBook},
using the Lyapunov function $V(q)$,
we can show that $\bk(q)$ is positive recurrent if \eqref{EQ:AD}
holds.

In \eqref{EQ:KRK}, since the next RB is uniformly chosen
at random by an active device
experiencing collision, it can be shown that
\be
\uE[\cR_{m,l} ([K_l (q)]_D)\,|\, [K_l (q)]_D = i]
= \frac{i}{M}.
	\label{EQ:iM}
\ee
Suppose that $K_m \le D$. 
Then, $[K_m]_D$ can be expressed as
\be
[K_m]_D  = K_m  - \sum_{n=1}^N \indicator (X_n^{K_m} = 1),
	\label{EQ:RKK}
\ee
where $X_n^{K_m}$ denotes the number of active devices
that choose SC $n$ when there are $K_m$ active
devices in RB $m$.
From \eqref{EQ:iM} and \eqref{EQ:RKK},
we have
\be
\uE[\cR_{m,l} ([K_l (q)]_D)\,|\, K_l (q) = K_l] =
\frac{K_l - \sum_{n=1}^N \Pr(X_n^{K_l} = 1)}{M}.
	\label{EQ:RPM}
\ee
Since the event of $X_n^{K_l} = 1$
means that there is only one device choosing
SC $n$ and the other devices choosing the other SCs
when there are $K_l$ active devices in RB $l$,
it can be shown that
\begin{align}
\Pr(X_n^{K_l} = 1)
 = \frac{K_l}{N} \left(1 - \frac{1}{N} \right)^{K_l-1}.
	\label{EQ:PX}
\end{align}
Substituting \eqref{EQ:PX} into \eqref{EQ:RPM}, and from
\eqref{EQ:AD}, 
it can be shown that
\begin{align}
& \uE[K_m(q+1)\,|\, \bk(q)] \cr
& = \uE
\left[ \sum_{l=1}^M \cR_{m,l} ([K_l(q)]_D) + 
\tilde A_m (q+1) \,|\, \bk(q)
\right] \cr
& =  \frac{1}{M} \sum_{l=1}^M U_D ( K_l (q) ) + 
\bar A_m \indicator(K_m (q) \le \bar K),
\end{align}
where
\be
U_D (K_l) = 
\left\{
\begin{array}{ll}
K_l, & \mbox{if $K_l > D$;} \cr
K_l \left(1 - \left( 1 - \frac{1} {N} \right)^{K_l-1} \right)
& \mbox{o.w.} \cr
\end{array}
\right.
\ee

Let
$\cC = \{\bk\,|\, K_m <  B_{D,N} \ \mbox{or} \ D < K_m \le M \bar K,
\ m = 1,\ldots, M\}$.
Suppose that $\bk (q) \in \Omega - \cC$.
In this case, 
we have $B_{D,N} \le K_m(q) \le D$.
Since
\begin{align*}
U_D (K_l(q)) & = 
K_l (q) \left(1 - \left( 1 - \frac{1} {N} \right)^{K_l (q)-1} \right) \cr
\bar A_m \indicator(K_m (q) \le \bar K) & = \bar A_m,
\end{align*}
the drift of $V(q)$ can be found as
\begin{align}
& \uE[V(q+1)\,|\, \bk(q)] - V(q) \cr
& = \sum_{m=1}^M U_D (K_m(q)) - K_m (q) + \bar A_m \cr
& = \sum_{m=1}^M \bar A_m - K_m (q) 
 \left( 1 - \frac{1} {N} \right)^{K_m (q)-1}.
\end{align}
Consider the function of $k$, $g(k) = k \left(1 - \frac{1}{N} \right)^{k-1}$,
which is an increasing function of $k$ when 
$0 \le k \le - \frac{1}{\ln
\left( 1 - \frac{1}{N} \right)} < N$.
From this and noting that $D \le N$, we have
$$
K_m (q) 
 \left( 1 - \frac{1} {N} \right)^{K_m (q)-1}
\le D \left( 1 - \frac{1} {N} \right)^{D-1} = B_{D,N},
$$
which implies that
\be
\uE[V(q+1)\,|\, \bk(q)] - V(q) \le \sum_{m=1}^M (\bar A_m - B_{D,N}).
	\label{EQ:VV}
\ee
From \eqref{EQ:VV}, if \eqref{EQ:AD} holds, we can conclude that
\be
\uE[V(q+1)\,|\, \bk(q)] - V(q) < 0,
\ \mbox{if} \ \bk(q) \in \Omega - \cC.
	\label{EQ:C1}
\ee
On the other hand, if $\bk (q) \in \cC$,
we have either
$ U_D(K_l(q)) - K_l(q) + \bar A_m 
\indicator(K_m (q) \le \bar K)
\le \bar A_m
$ (when $K_l (q) > D$) or 
$U_D(K_l(q)) - K_l(q) + \bar A_m
\le \bar A_m$
(when $K_l (q) < B_{D,N} $).
Thus, we have
\be
\uE[V(q+1)\,|\, \bk(q)] - V(q) \le \sum_{m=1}^M \bar A_m,
\ \mbox{if} \ \bk(q) \in \cC.
	\label{EQ:C2}
\ee
From \eqref{EQ:C1} and \eqref{EQ:C2},
we can see that $\bk(q)$ is positive recurrent
under the Foster-Lyapunov stability criterion
\cite{HajekBook}.

\section{Proof of Theorem~\ref{T:X}}	\label{A:X}
Under {\bf A1} and {\bf A2},
from \eqref{EQ:UKm},
it can be shown that
$$
\uE \left[ [K_m (q)]_D \right]
=  \sum_{n=0}^D \left(n - 
n \left( 1 - \frac{1}{N} \right)^{n-1} \right) p_n 
+ \sum_{n=D+1}^\infty n p_n,
$$
where $p_n = \Pr(K_m (q)= n)$.
Since $K_m (q) \sim {\rm Pois} (\lambda_1)$,
we have
\begin{align}
\uE \left[ [K_m (q)]_D \right]
& = \lambda_1 - \lambda_1 \sum_{n=0}^{D-1}
n \left( 1 - \frac{1}{N} \right)^{n} 
\frac{\lambda_1^n e^{-\lambda_1}}{n!} \cr
& = \lambda_1
\left(
1 - e^{- \frac{\lambda_1}{N}}
\frac{\Gamma(D, \nu_1)}{(D-1)!} \right),
\end{align}
which completes the proof.
\section{Proof of Theorem~\ref{T:2}}	\label{A:T2}

For convenience, let $\bar \lambda_1^*$ represent
the solution to \eqref{EQ:llmax}.
In \cite{Choi17_Stability},
it is shown that
$\bar \lambda_1^*$ is smaller than $D$ for $D > 1$,
i.e., $\bar \lambda_1^* < D$.
We can also show that
\be
\bar \lambda_{\rm max} < \bar \lambda_1^* < D, \ \mbox{for} \ D > 1.
	\label{EQ:llD}
\ee
The first inequality in
\eqref{EQ:llD} is obtained by the fact that
$\bar \lambda_1 e^{-\bar \lambda_1
(1 - 1/N)} \sum_{n=0}^{D-1}\frac{(\bar \lambda_1(1-1/N) )^n}{n!}
< \bar \lambda_1$ from \eqref{EQ:llmax}.
Thus, there exists 
$\epsilon > 0$, which is independent of $N$,
such that
$\bar \lambda_{\rm max} = D- \epsilon$.
For given $D$, there exists a $N_*$ such that the following
holds for $N \ge N_*$:
\begin{align*}
B_{D,N} - \bar \lambda_{\rm max} 
& = D \left(1 - \frac{1}{N} \right)^{D-1} - D + \epsilon \cr
& \ge D \left( 1 - \frac{D-1}{N} - 1 + \frac{\epsilon}{D} \right) \cr
& = D \left(  \frac{\epsilon}{D} - \frac{D-1}{N} \right) > 0,
\end{align*}
which completes the proof.

\bibliographystyle{ieeetr}
\bibliography{cs}
\end{document}